\documentclass[preprint,showpacs,showkeys,preprintnumbers,amsmath,amssymb,superscriptaddress]{revtex4}

\usepackage{graphicx}
\usepackage{dcolumn}
\usepackage{bm}
\usepackage{multirow}
\usepackage{mathtools}
\usepackage{amsmath}
\usepackage{array}
\usepackage{xcolor}
\newcolumntype{P}[1]{>{\centering\arraybackslash}p{#1}}

\begin{document}

\preprint{}

\title{Majority-vote model with limited visibility: An investigation into filter bubbles}

\author{Andr\'e L. M. Vilela}
 \email{andre.vilela@upe.br}
\affiliation{F\'isica de Materiais, Universidade de Pernambuco, Recife, PE 50720-001, Brazil}
\affiliation{Center for Polymer Studies and Department of Physics, Boston University, Boston, MA 02215, USA}
  
\author{Luiz Felipe C. Pereira}
\email{pereira@df.ufpe.br}
\affiliation{Departamento de F\'{\i}sica, Universidade Federal do Rio Grande do Norte, Natal, RN 59078-970, Brazil}
\affiliation{Departamento de F\'{\i}sica, Universidade Federal de Pernambuco, Recife, PE 50670-901, Brazil}

\author{Laercio Dias}
\affiliation{Departamento de F\'{\i}sica, Universidade Federal do Rio Grande do Norte, Natal, RN 59078-970, Brazil}

\author{H. Eugene Stanley}%
\affiliation{Center for Polymer Studies and Department of Physics, Boston University, Boston, MA 02215, USA} 
  
\author{Luciano R. da Silva}
\affiliation{Departamento de F\'{\i}sica, Universidade Federal do Rio Grande do Norte, Natal, RN 59078-970, Brazil}
\affiliation{National Institute of Science and Technology for Complex Systems, Centro Brasileiro de Pesquisas Fsicas, Rio de Janeiro, RJ 22290-180, Brazil}

\date{\today}

\begin{abstract}

The dynamics of opinion formation in a society is a complex phenomenon where many variables play an important role. 
Recently, the influence of algorithms to filter which content is fed to social networks users has come under scrutiny.
Supposedly, the algorithms promote marketing strategies, but can also facilitate the formation of filters bubbles in which a user is most likely exposed to opinions that conform to their own.
In the two-state majority-vote model an individual adopts an opinion contrary to the majority of its neighbors with probability $q$, defined as the noise parameter.
Here, we introduce a visibility parameter $V$ in the dynamics of the majority-vote model, which equals the probability of an individual ignoring the opinion of each one of its neighbors.
For $V=0.5$ each individual will, on average, ignore the opinion of half of its neighboring nodes.
We employ Monte Carlo simulations to calculate the critical noise parameter as a function of the visibility $q_c(V)$ and obtain the phase diagram of the model. 
We find that the critical noise is an increasing function of the visibility parameter, such that a lower value of $V$ favors dissensus.  
Via finite-size scaling analysis we obtain the critical exponents of the model, which are visibility-independent, and show that the model  belongs to the Ising universality class. 
We compare our results to the case of a network submitted to a static site dilution, and find that the limited visibility model is a more subtle way of inducing opinion polarization in a social network.

\end{abstract}



\maketitle

\section{INTRODUCTION}

Opinion formation in a real society is a complex phenomenon where many variables have a marked influence. 
For simplicity, consider a situation where a population is called upon to choose between two options, such as an election in a bipartisan system or a referendum.
Even in this simple scenario, several factors can influence the decision making process of each individual.
One might consider aspects such as
herd mentality and social pressure \cite{Asch1955, Cont2000}, conformity \cite{Milgram1961}, confirmation bias \cite{Nickerson1998}, social dominance \cite{Sidanius2001, Cook2014}, and social influence bias \cite{Muchnik2013, Walter2014} among others. 
Recently, one more aspect has become particularly relevant in opinion formation processes, the influence of individuals and corporations via social networks \cite{Grabowski2009, Effing2011, Xiong2014}.

In general, social platforms employ specific algorithms to filter which content is fed to a given user based on the user's interactions in the digital world.
One of the main objectives of such algorithms is to promote target-oriented marketing strategies.
Nonetheless, the same algorithms can also promote the formation of so-called filters bubbles \cite{Flaxman2016, Geschke2019}, in which a user is more likely to be exposed to news and opinions which conform to their current beliefs.
This kind of filter can isolate users in bubbles, hence the name, where a single opinion is prevalent, even if that opinion is unpopular among the total population.
Notice that the formation of filter bubbles depends on the presence of individuals on each others news feed, which is preferred if both have similar views and opinions.

Modeling opinion formation using statistical physics methods is not a simple task, even if the objective is only a minimally acceptable representation of social reality \cite{Galam1997, Galam1999, Fernandez-Gracia2014}.
Nonetheless, many advances have been made in the field of opinion dynamics, sometimes referred to as Sociophysics \cite{Castellano2009, Galam2012, Kutner2019}.
Several opinion formation models proposed by physicists are based on the description employed for magnetic systems, such as the Ising model \cite{DaSilva1991, Galam1997, Sznajd2000, Li2019}.
In this analogy, magnetic moments (spins) become individuals and the exchange energy becomes the social interaction.
Therefore, each individual holds one of several possible opinions, corresponding to discrete spin states, and the opinion of each individual is influenced by the opinion of its neighbors.
The atomic lattice becomes the network of social interactions, where each node is occupied by an individual agent, and its topology can take the form of regular or complex networks.

Arguably, two of the most popular models in opinion dynamics are the voter model \cite{Clifford1973, Liggett2005} and its generalization, known as the majority-vote model \cite{Oliveira1992}.
Originally proposed by Oliveira, the majority-vote model adds a noise parameter to the voter dynamics, which corresponds to the probability of an individual adopting an opinion that differs from the majority of its neighbors \cite{Oliveira1992, Oliveira1993}.
The noise parameter is analogous to the temperature in the Ising model, and sometimes it is dubbed the social temperature.
Starting from zero noise and increasing its value the system undergoes a phase transition from an ordered phase, where one opinion prevails, to a disordered phase, with no dominant opinion which corresponds to a polarized society.
The critical noise at which the phase transition takes place is analogous to the critical temperature in the Ising model.

At first, the two-state majority-vote model was investigated on a regular square lattice.
It was later generalized to a cubic lattice \cite{Acuna2012}, small-world networks \cite{Campos2003, Medeiros2005}, random graphs \cite{Felipe2005, Lima2008},  scale-free networks \cite{Lima2006, Lima2019}, and spatially embedded networks \cite{Cesar2016}.
The impact of site dilution and agent diffusion on the critical behavior of the majority-vote model has also been addressed recently \cite{Nuno2012, Nuno2012Err}, as well as the presence of two types of noises \cite{Nuno2016}.
The model has been generalized to three-states and investigated on a square lattice \cite{Brunstein1999, Tome2002, Lima2012}, random graphs \cite{Melo2010}, and scale-free networks \cite{Vilela2020}.
Further generalizations include a continuous-state version of the model  \cite{Adauto2005}, the majority-vote model with inertia \cite{Fiore2018, Fiore2018a}, and the presence of anticonformists \cite{Andrzej2018} and strong opinions \cite{Vilela2018}.
A variation of the majority-vote model,  where a set of individuals tries to influence the opinion of their neighbouring counterparts has been named the block-voter model \cite{Cesar2011, Araujo2018}.

In this study, we introduce a probabilistic visibility parameter in the majority-vote model in order to mimic the filter bubble effect.
When the visibility is smaller than unit, the individual will ignore the opinion of some of its neighbors.
We perform Monte Carlo simulations of the two-state majority-vote model on a square lattice to obtain the critical social temperature as a function of the visibility, i.e. the phase diagram of the model. 
Furthermore, we employ finite-size scaling techniques to estimate the critical exponents which characterize the phase transition observed in this model.
Our results show that the critical noise is an increasing function of the visibility, and as we lower the visibility a smaller amount of noise is required to polarize the system.
The critical exponents equal those of the Ising model, which places the majority vote model with limited visibility in the Ising universality class, in accordance with the conjecture by Grinstein et al. \cite{Yuhe1985}.

\section{Limited-visibility MODEL}

The majority-vote model describes a system of interacting individuals where each one is allowed to be in one two possible states $\sigma_i = \pm 1$.
This two-state model evolves in time following a simple update rule: each individual assumes the state of the majority of its neighbors with probability $1-q$ and the opposite state with probability $q$, independent of its previous state.
It is also possible to express the probability of a given individual changing their opinion in terms of their current state and that of the majority of its neighbors \cite{Oliveira1992}.
The system presents an orderâdisorder phase transition as the noise parameter $q$ reaches a critical value $q_c$. 

The majority-vote model with limited visibility aims to include in the opinion dynamics a parameter that quantifies how each individual $i$, perceives its neighborhood $\Lambda_i$ or is isolated from it. 
In this model, we propose that each individual is influenced by the opinion of its visible neighborhood $\Lambda_i^{*}$, governed by a  probability $V_i$. 
The parameter $V_i$ is the visibility, defined in the range $0 \leq V_i \leq 1$.
If $V_i=1$ the individual considers the opinion of all of its neighbors equally, while for $V_i=0$ the individual ignores the opinion of all the neighbors.
For $0 < V_i < 1$, the opinion of each neighbor is considered with probability $V_i$. 
Hence, in the case of $V_i = 1/2$, the individual will on average consider the opinion of half of its neighbors, ignoring the opinion of the other half. 
Thus, $V_i$ effectively stands for the sight range of a given individual.

In this work, we consider a single visibility parameter for all individuals in the network $V_i = V$ for all $i$.
Such dynamics simulate in a simplified way the effect of filters in social networks. 
Because the algorithm supposedly chooses the content presented to an individual according to their reactions to previous content, news from a certain neighbor may be visible on a given day  but not on subsequent days. 
We model such behavior as a simple random choice of which content will be seen at any given moment, abstracting details from the real implementation of recommendation algorithms, without loss of generality.

We define the visibility index $I(V)$ as
\begin{equation}
I(V) =
\begin{cases}
 1, & \text{with probability } V; \\ 
 0, & \text{with probability } 1-V,
\end{cases}
\label{vparameter}
\end{equation}
where $V$ is the visibility parameter.
The dynamics of the model with limited visibility is similar to that of the standard majority-vote model.
The opinion of an individual $\sigma_i$ is flipped with probability
\begin{equation}
w(\sigma_i) = \frac{1}{2}\left\{1-(1-2q) \, \sigma_i \, \textrm{sgn}\left[\sum_{\delta = 1}^{k_i}I(V) \, \sigma_{i+\delta}\right]\right\},
\label{probw}
\end{equation}
where $\textrm{sgn}(x) = +1, 0, -1$ for $x < 0$, $x = 0$, and $x > 0$, respectively. 
The sum runs over all $k_i$ neighbors connected to the individual $i$. 
In our model $I(V)$ acts dynamically and it is tested for each neighbor.
For the special case where the selected individual cannot interact with any of its neighbors in a given trial, its opinion remains unchanged.
In other words, if $I(V) = 0$ for all $k_i \in \Lambda_i$ (or $\Lambda_i^{*} = \emptyset$), then $w(\sigma_i) = 0$. 
Here, we place the individuals on the nodes of a regular two-dimensional square lattice of size $N = L \times L$, such that $k_i = 4$ for all $i \in N$. Naturally, for $V = 1$ we recover the original majority-vote model \cite{Oliveira1992}. 


Before moving on to the simulation details, we must remark one detail concerning the nature of the visibility parameter.
At first sight, one might think that the limited  visibility introduced in our model is equivalent to a standard site dilution \cite{Reimer1994, Martins2007}.
However, this is certainly not the case.
The dilution is a static feature of the network since an inactive site will remain dormant during the dynamics of the system.
Meanwhile, the visibility of an individual's neighborhood proposed here is a fully dynamic feature, and it may change at each time step.
In a sense, the visibility parameter plays the role of a dynamic dilution.
Therefore, we do not expect our results to agree with previous studies on statically diluted systems. 

\section{Simulation Details}

In order to investigate the effect of the  visibility in the majority-vote model we define an order parameter, analogous to the magnetization per spin in the Ising model, given by 
\begin{equation}\label{averageopinion}
    m = \frac{1}{L^{2}}
    \left|\sum_{i=1}^{L^{2}} \sigma_i\right|.
\end{equation}
We also consider the average of the order parameter given by 
\begin{equation}\label{magnetization}
    M(q, V, L)= \left\langle\left\langle
    m\right\rangle_{t}\right\rangle_{c},
\end{equation}
where $ \langle . . . \rangle_t $ stands for time averages taken in the stationary regime, and $ \langle . . . \rangle_c$ indicates   configurational averages taken over independent realizations.
The behavior of the model is further characterized by the scaled variance of $m$, which is an extensive quantity analogous to the magnetic susceptibility
\begin{equation}\label{chi}
    \chi(q, V, L)= L^{2} \left[ \langle\langle m^2\rangle_{t}\rangle_{c}-{\langle\langle m\rangle_{t}\rangle_{c}^2} \right],
\end{equation}
and the kurtosis of $m$, in the form of the Binder fourth-order cumulant \cite{Binder1981}
\begin{equation}\label{bindercumul}
    U(q, V, L)= 1-\frac{\langle\langle m^4\rangle_{t}\rangle_{c}}{3{\langle\langle m^2\rangle_{t}\rangle_{c}^2}}.
\end{equation}
All of the above quantities depend on the noise parameter $q$, the visibility $V$, and the system size $L$.
Therefore, we expect our model to present a phase transition as $q$ and $V$ are varied, and thus characterize its critical behavior by finite-size scaling analysis \cite{Stanley1971}.

In the critical region around the phase transition, the susceptibility presents a maximum at a size-dependent pseudocritical noise parameter, which is related to the true critical noise by
\begin{equation}\label{finiteq}
        q_{c}(V, L) = q_c(V) + rL^{-1/\nu},
\end{equation}
where $r$ is a numerical constant, and $\nu$ is one of the critical exponents.
The behavior of $M$ and $\chi$ in the vicinity of the phase transition is defined by the following finite-size scaling relations %
\begin{equation}\label{finitemag}
    M(q, V, L) = L^{-\beta/ \nu}\widetilde{M}(\epsilon L^{1/\nu}),
\end{equation}
\begin{equation}\label{finitequi}
        \chi(q, V, L) = L^{\gamma/\nu}\widetilde{\chi}(\epsilon L^{1/ \nu}),
\end{equation}
where $\beta/\nu$ and $\gamma/\nu$ are  critical exponent ratios, and $\epsilon = q - q_c$ is the distance to the critical noise. 
For each value of $V$, $\widetilde{M}$ and $\widetilde{\chi}$ are universal scaling functions of the variable  $x = \epsilon L^{1/\nu}$.
The exponent ratios are related to the system dimension by the so-called  hyperscaling relation \cite{Stanley1971} 
\begin{equation}
\label{hyperscaling}
2\beta/\nu + \gamma/\nu = d, 
\end{equation}
which in our case should be satisfied with $d = 2$, since our model is defined on the square lattice \cite{Vilela2020}.
The fourth-order cumulant presents its own scaling in terms of a universal function
\begin{equation}\label{finitecumul}
        U(q, V, L) = \widetilde{U}(\epsilon L^{1/ \nu}).
\end{equation}
However, unlike $M$ and $\chi$, the cumulant assumes a size-independent universal value at the critical point $\widetilde{U}(0)$, whose numerical value can be used to identify the university class of the system \cite{Binder1981, Malakis2014}.

We perform Monte Carlo simulations of the majority-vote model with limited visibility on  square lattices with periodic boundary conditions and linear sizes ranging from $L = 8$ to $200$. 
In our simulations, time is measured in Monte Carlo steps (MCS), which corresponds to $N$ attempts of changing the state of the individuals in the network. 
Each simulation begins in a fully ordered state, with all spins set to $\sigma_i(t = 0) = +1$ for all $i \in N$, and thus $m=1$. 
For any finite $q$, the system will take a certain number of MCS to reach its steady state, so we discard the first $5 \times 10^4$ MCS in each simulation.
Time averages were calculated over the next $7 \times 10^5$ MCS, and up to $100$ independent samples were considered to calculate configurational averages.

\section{NUMERICAL RESULTS}

\begin{figure}[htb]
\centering
\includegraphics[width=1.0\linewidth]{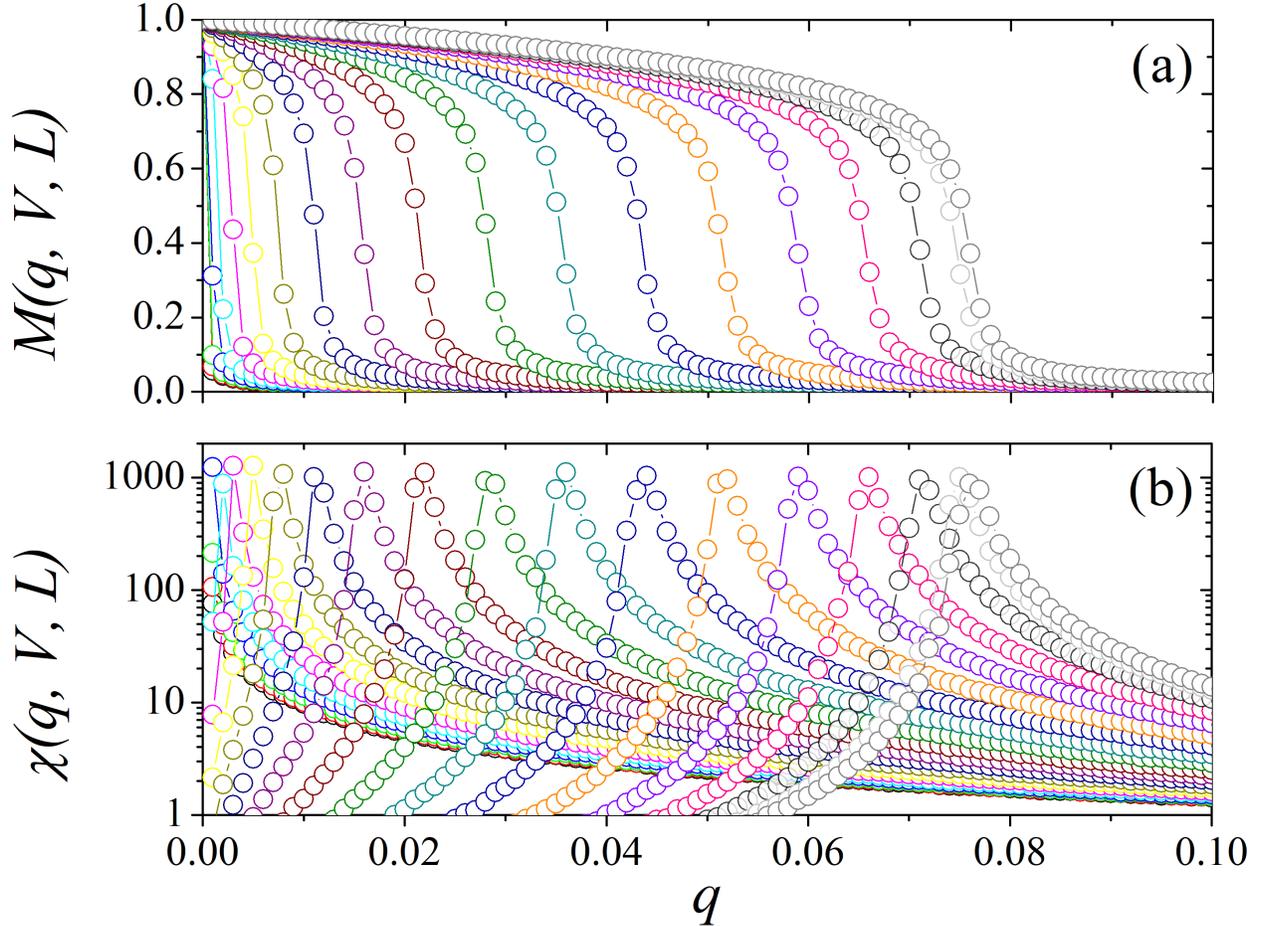}
\caption{(Color online) The social temperature dependence of (a) order parameter and (b)  susceptibility for the majority-vote model with limited visibility for $L = 200$. 
The visibility parameter increases from left $V=0.05$ to right $V=1$ with $\Delta V = 0.05$. Error bars are smaller than the symbol size. 
Lines are guides to the eye.}
\label{mxv}
\end{figure}

Fig. \ref{mxv} presents the dependence of the average order parameter and the corresponding susceptibility on the social temperature $q$, for the majority-vote model with limited visibility on a square lattice of size $L=200$.
Each curve corresponds to a visibility parameter ranging from $0.05$ to $1.00$ (from left to right) with increments $\Delta V = 0.05$. 
For each value of $V$, the system undergoes a phase transition from consensus to dissensus as the noise parameter increases. 
The curves show that as the visibility decreases the value of $q$ at which the system becomes disordered also decreases.

In the two-state majority-vote model, consensus is a macroscopic state where one the two opinions is adopted by the majority of the society, similar to a ferromagnetic state of a magnet. 
The dissensus represents a macroscopic behavior that mimics the paramagnetic state, where the two opinions are distributed symmetrically and $M \sim 0$. 
The pseudocritical noise values $q_c(V, L)$ are located at the peaks of the susceptibility, which are in agreement with the behavior of $M(q, V, L)$. 
From the data in Fig. \ref{mxv}, we conclude that decreasing the visibility decreases the critical noise required to reach a disordered state. 
This behavior indicates that a limited visibility weakens consensus in the system, since a smaller amount of noise is required for the phase transition to take place.

\begin{figure*}[htb]
\centering
\includegraphics[width=1.0\linewidth]{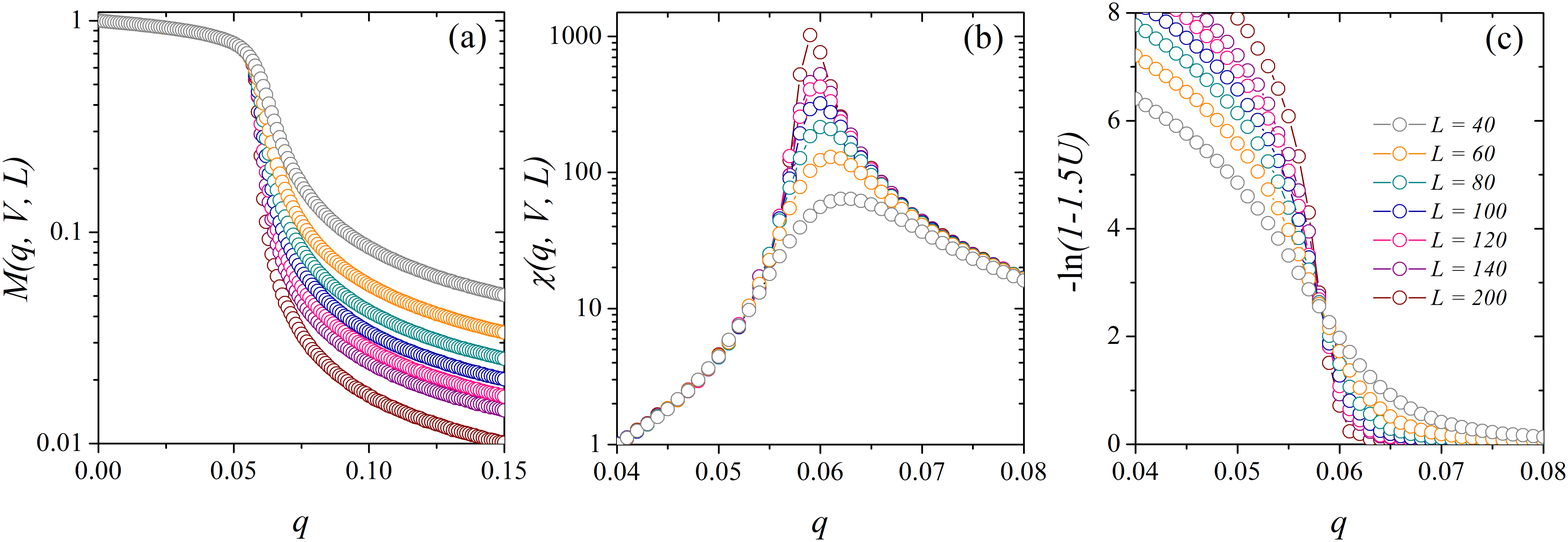}
\caption{(Color online) (a) Order parameter, (b) susceptibility, and (c) re-scaled fourth-order cumulant as a function of the noise parameter for several values of the system size $L$ and fixed visibility $V = 0.8$. 
Within the accuracy of the data, all curves in (c) intersect at $q_c = 0.0582(1)$. 
Lines are just a guide to the eyes.}
\label{varLv080}
\end{figure*}

In Fig. \ref{varLv080} we fix the visibility $V = 0.8$ and plot the average order parameter, the susceptibility, and the re-scaled fourth-order cumulant as functions of $q$ for lattices with  sizes $L = 40, 60, 80, 100, 120, 140$ and $200$. 
Fig. \ref{varLv080}(a) shows a phase transition from an ordered state to a disordered one as the social temperature goes above some critical value of $q$. 
In Fig. \ref{varLv080}(b), the susceptibility  reaches a maximum in the critical region $q \simeq q_c(V, L)$, which become sharper for larger system sizes. 
In Fig. \ref{varLv080}(c) the cumulant for systems of different size intersect at the critical point $q_c(V=0.8) = 0.0582(1)$. 
Once again, the results in Fig. \ref{varLv080}(c) indicate that a limited visibility lowers the critical value of the social temperature $q$ compared to the isotropic case, since $q_c(V=1) = 0.075(1)$ \cite{Oliveira1992, Oliveira1993}. 

\begin{figure}[htb]
\centering
\includegraphics[width=0.8\linewidth]{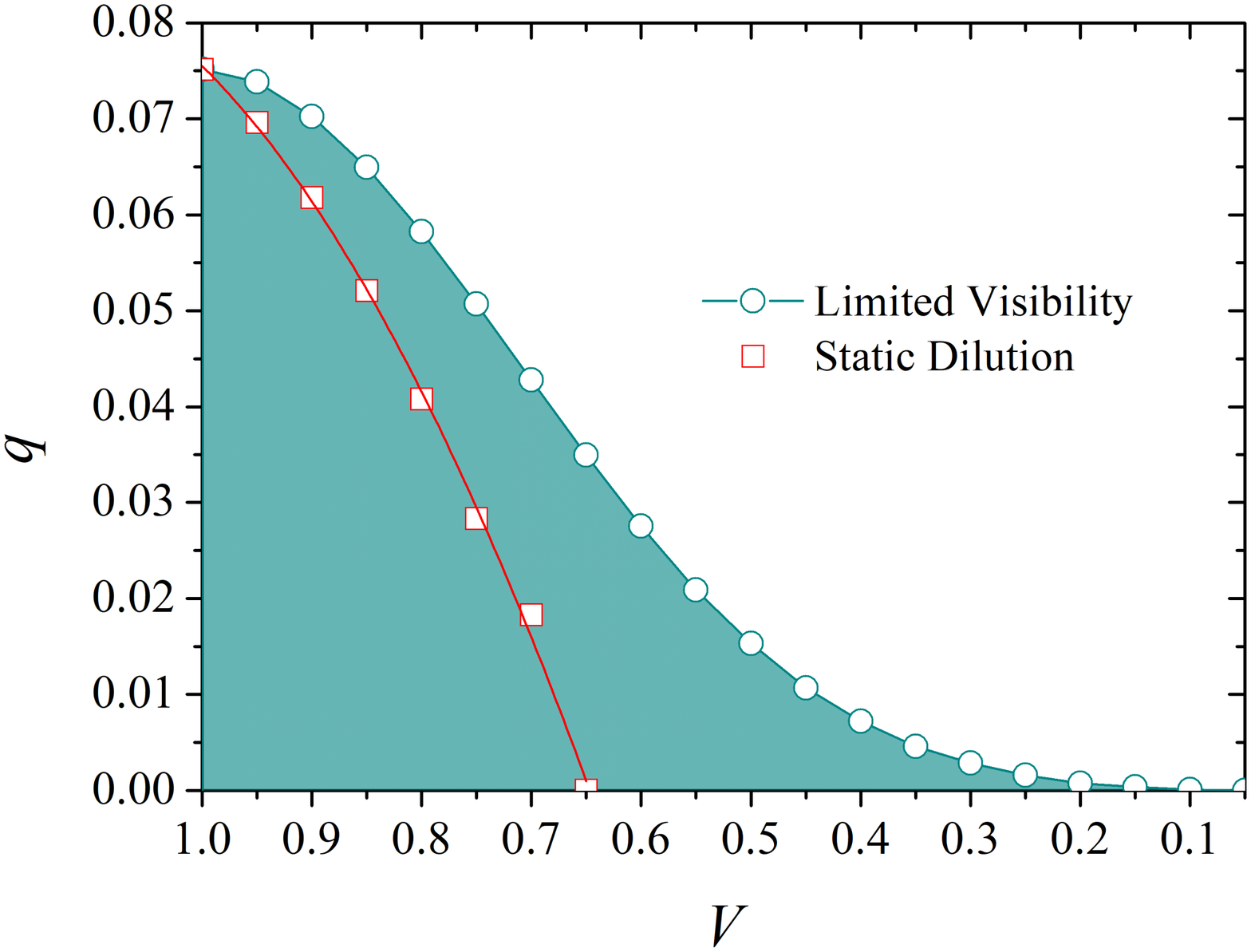}
\caption{(Color online) Phase diagram of the majority-vote model on a square lattice, obtained from the crossing of the Binder cumulant. 
Circles represent the critical noise for the  model with limited visibility, and squares are for the static site dilution case.
Error bars are smaller than the symbols. 
In the green (white) region, the system presents consensus (disensus) for the model with limited visibility. 
The the red line is a polynomial fit to the data $q_c(V) = a + bV+ cV
^2$, where $a = -0.32(3)$, $b = 0.69(7)$, and $c = -0.29(4)$.}
\label{pdiagram}
\end{figure}

The dependence of $q_c$ on $V$ is shown in Fig. \ref{pdiagram}, where we present the phase diagram of the majority vote model with limited visibility. 
In this plot, we consider the critical points $q_c(V)$ estimated from the crossing of the Binder cumulant for each $V$. 
The circles represent the numerical values and the error bars are smaller than the symbols.
The critical points separate an ordered phase, for which the system presents consensus, from a disordered phase, where dissensus dominates. 
Once again, we observe that the lower the visibility of an individual's neighborhood, the lower the critical noise that drives the system to the disordered state. 
Thus, the critical social temperature is an increasing function of the visibility parameter, and limiting the visibility can effectively hinder the formation of an ordered state.

\begin{table}[htb]
\centering
\caption{\label{tab1} The critical noise as a function of the visibility parameter for our model $q_c(V)$ and for the static dilution case.} 
\begin{ruledtabular}
\begin{tabular}{ccc}
$V$ & $q_c(V)$ & $q_c^{\textrm{static dilution}}$ \\
\hline
$1.0$ & $0.0751(1)$  & $0.0751(1)$ \\
$0.9$ & $0.0703(1)$  & $0.0618(1)$ \\
$0.8$ & $0.0582(1)$  & $0.0408(5)$ \\
$0.7$ & $0.0428(1)$  & $0.0183(5)$ \\
$0.6$ & $0.0275(1)$  & $-$ \\
$0.5$ & $0.0153(1)$  & $-$ \\
$0.4$ & $0.0072(1)$  & $-$ \\
$0.3$ & $0.0029(1)$  & $-$ \\
$0.2$ & $0.0007(1)$  & $-$ \\
$0.1$ & $0.00009(5)$ & $-$ \\
\end{tabular}
\end{ruledtabular}
\end{table}

Table \ref{tab1} lists the values of the critical noise for several values of the visibility parameter $V$.
The table also presents results for the static dilution case which will be discussed later.
The results for $q_c(V)$ indicate that the control of the visibility is the key to promote or suppress polarization in a society.
Therefore, bubble filters have a negative impact in the formation of consensus in a population, since individuals are more likely to ignore opinions opposite to their own.
We note that for $V < 0.5$, the amount of social noise required to destroy consensus is lower than $0.01$. 
Therefore, if we consider a social network where individuals agree with only half of its connections, even the smallest amount of noise (or disagreement) is capable of originating a polarized society. 
In what follows, we focus our presentation for networks where $V \geq 0.5$.

\begin{figure}[htb]
\centering
\includegraphics[width=0.8\linewidth]{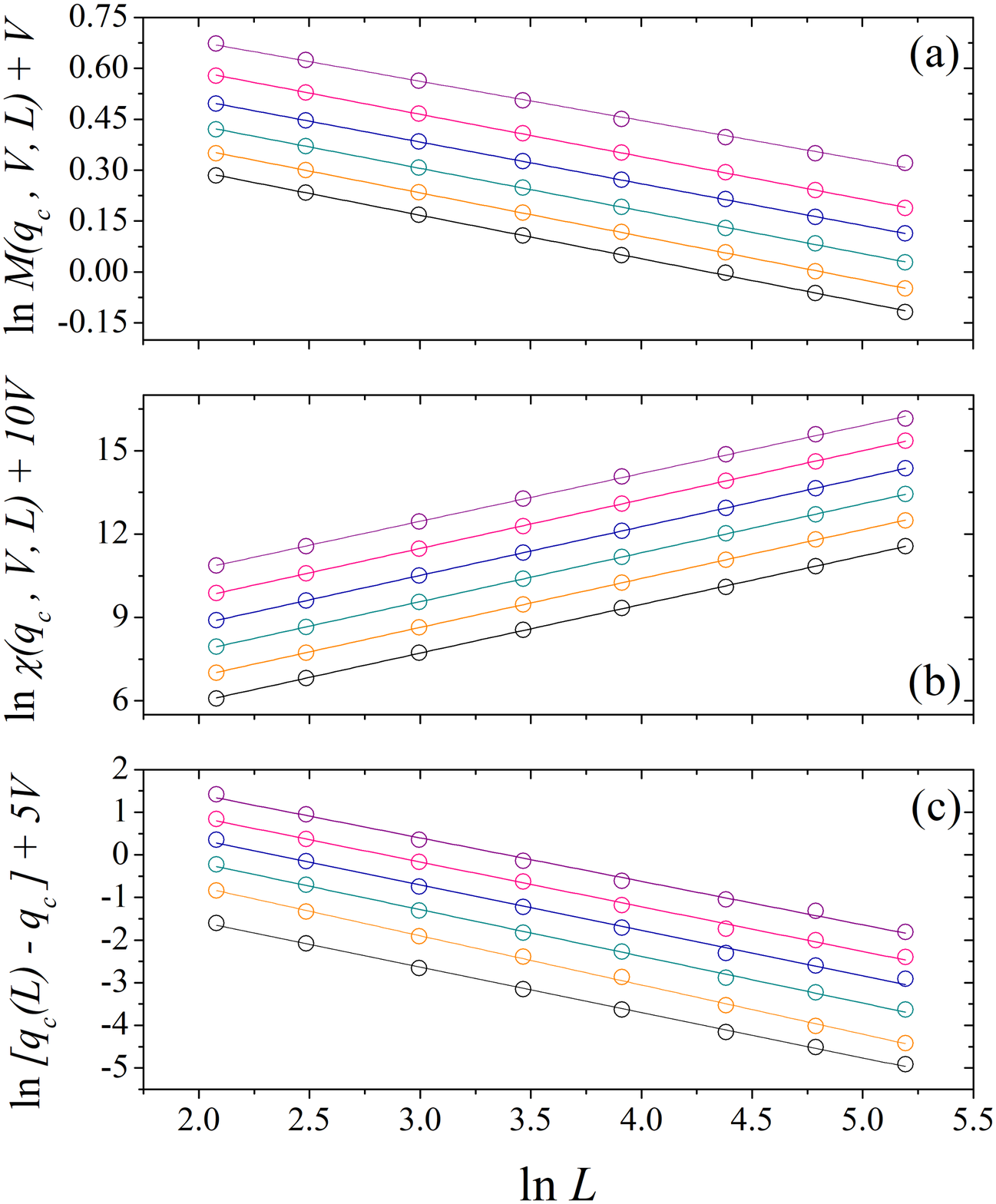}
\caption{(Color online) Plot of (a) $\ln M(q_c, V, L)$, (b) $\ln \chi(q_c, V, L)$ and $\ln [q_c(V, L) - q_c(V)]$ versus $\ln L$. From top to bottom, $V = 1.0, 0.9, 0.8, 0.7, 0.6$ and $0.5$.
Straight lines are obtained from a linear regression to the data, and their slope equals the respective critical exponents. 
Taking the error bars into consideration, we find (a) $\beta/\nu = 0.125$, (b) $\gamma/\nu = 1.75$, and (c) $1/\nu = 1.00$.}
\label{lnmxq}
\end{figure}

The two upper panels of Fig. \ref{lnmxq} show (a) $\ln M$ and (b) $\ln \chi$ at the critical point $q = q_c(V)$ versus $\ln L$ for $V = 1.0, 0.9, 0.8, 0.7, 0.6$ and $0.5$.
The angular coefficient of the straight lines, obtained from a linear regression to the simulation data, confirm the scaling of $M$ and $\chi$ according to Eqs. (\ref{finitemag}) and (\ref{finitequi}). 
In Fig. \ref{lnmxq}(c) we  plot $\ln [q_c(V, L) - q_c(V)]$ versus $\ln L$, from which we obtain the critical exponent $1/\nu$ according to Eq. (\ref{finiteq}). 
The slope of the lines for Fig. \ref{lnmxq} (a), (b) and (c) correspond to critical exponent ratios $\beta/\nu = 0.125$, $\gamma/\nu = 1.75$ and $1/\nu = 1.0$.
Those values indicate that the majority-vote model on a square lattice with limited visibility  belongs to the Ising universality class. 
This finding is further corroborated by the  size-independent universal value calculated for the Binder cumulant $\widetilde{U}(0)=0.611(2)$.
Finally, we note that the hyperscaling relation $2\beta/\nu + \gamma/\nu = d$ is satisfied with $d = 2$, as one would expect for the square lattice \cite{Stanley1971, Vilela2020}.

\begin{figure*}[htb]
\centering
\includegraphics[width=1.0\linewidth]{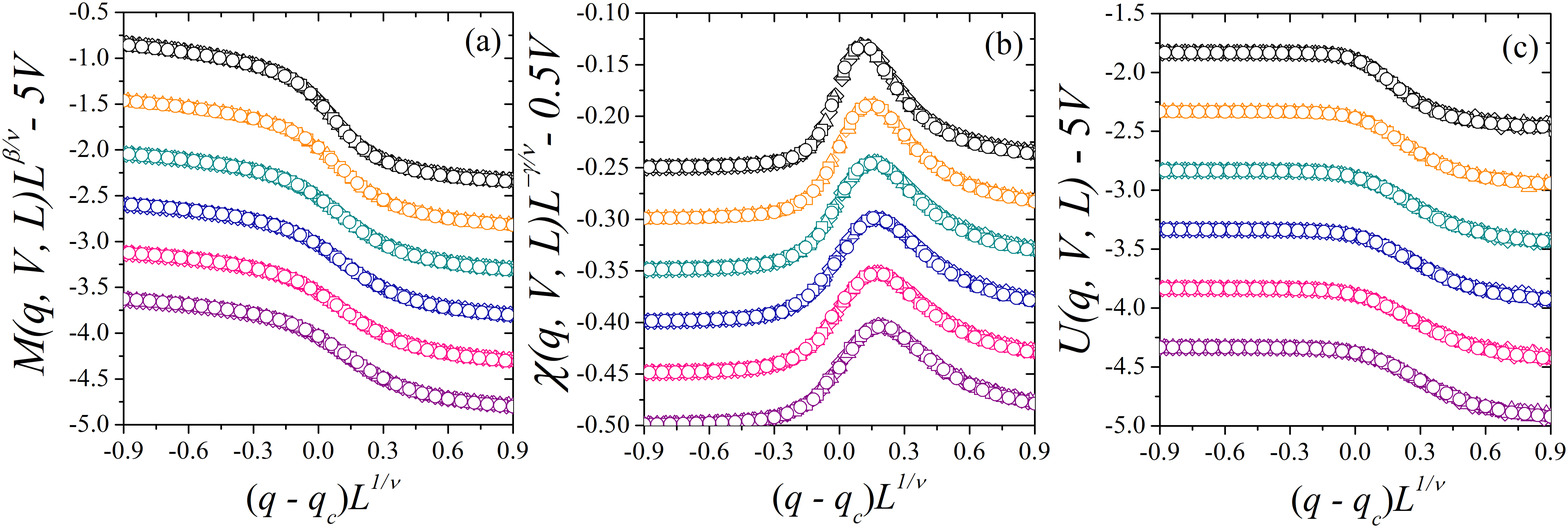}
\caption{(Color online) Scaling plot of the (a) order parameter, (b) susceptibility and (c) fourth-order cumulant for lattice sizes $L = 60, 80, 100$ and $120$. 
From top to bottom we have $V = 0.5, 0.6, 0.7, 0.8, 0.9$ and $1.0$. For all scaling plots, we use $\beta/\nu = 0.125$, $\gamma/\nu =1.75$ and $1/\nu = 1.0$.}
\label{allcollapse}
\end{figure*}

Using the finite size scaling relations from Eqs. (\ref{finitemag})-(\ref{finiteq}) with the critical exponents of the model it is possible to obtain the universal functions $\widetilde{M}(x)$, $\widetilde{\chi}(x)$ and $\widetilde{U}(x)$ shown in Fig. \ref{allcollapse} versus the scaled variable $x = \epsilon L^{1/\nu}$. 
The curves are shifted down to avoid overlapping between them, but yield only one size-independent universal curve for $\widetilde{M}$ and $\widetilde{\chi}$, for each visibility.
We obtain similar results for the scaled $M$, $\chi$ and $U$ for all $V \in (0, 1]$, indicating that the majority-vote model with limited visibility on a square lattice indeed belongs to the Ising universality class. 
Furthermore, this result supports the conjecture stating that irreversible systems with up-down symmetry belong to the universality class of the equilibrium Ising model for regular lattices \cite{Yuhe1985}.

\section{Limited Visibility versus Static Dilution}

In Fig. \ref{pdiagram} and Tab. \ref{tab1} we also show the critical noise values for the majority-vote model with a static dilution (red squares in the phase diagram). 
In this case, for each realization, we select a fraction $V$ of nodes that remain present in the network, and permanently remove the remaining $1-V$ fraction.
For the sake of consistency with the limited visibility model, we enforce that an individual remains in its original state if there are no neighbors connected to it. 
The solid red line in Fig. \ref{pdiagram} was obtained from a polynomial fit to the data in the form $q_c(V) = a + b V + c V^2$, where $a = -0.32(3)$, $b = 0.69(7)$, and $c = -0.29(4)$. 
This result agrees well with previous investigations on the majority-vote dynamics with dilution \cite{Nuno2012, Nuno2012Err}.
%


The simulation results show that the critical noise required to reach the disordered state is lower for the static dilution when compared with the dynamic dilution of the limited visibility model. 
We conclude that the static dilution and the limited visibility model are both capable of promoting polarization, yet via different mechanisms. 
While the static dilution improves order undergoing an abrupt interference, where individuals may notice the permanent absence of their neighboring nodes, the dynamic dilution promotes ordering by a more subtle approach. 
In the latter, one individual will most likely interact with an ever-changing reduced neighborhood at any given time, instead of interacting exclusively with the same individuals.
We remark that the limited visibility model also presents a broader tuning range for the control parameter $V$, where one can improve the ordering by adjusting this value smoothly and continuously.

\section{CONCLUSION AND FINAL REMARKS}

In this work, we introduced a visibility parameter in the two-state majority-vote model in order to capture the possible effects of filter bubbles on opinion formation dynamics.
The visibility parameter, $0 \leq V \leq 1$, equals the probability of an individual considering the opinion of one of its neighbors when determining the majority opinion.
When the visibility equals unit we recover the original model.
The geometric structure of social interactions supports an order-disorder phase transition when the social temperature is increased above a critical value, which is an increasing function of the visibility parameter.
Our investigation shows that one might promote or suppress polarization in a social network by controlling the visible neighborhood of an individual, in other words, controlling the effective influence of its closest social group.

Numerical results exhibit the typical finite-size effects for the order parameter when the system undergoes a second order phase transition. 
We obtain the critical exponents of the model via finite-size scaling to find $\beta / \nu = 0.125$, $\gamma /\nu = 1.75$ and $1/\nu = 1.0$. 
Moreover, using the hyperscaling relation we obtain $d = 2$ as expected for a two-dimensional square lattice of social interactions. 
The calculated critical exponents do not depend on the visibility parameter and place the majority-vote model with limited visibility in the Ising universality class.
 
Our investigation covers a key feature on how to promote or suppress polarization in a social network. 
We remark that preventing consensus by means of a limited visibility might also be implemented to prevent negative effects, such as the spread of fake news.
A natural extension of the limited-visibility opinion dynamics include the use of complex networks to map the geometry of social interactions. 
We also avail that the use of different probability distributions for the visibility index $I(V)$ might produce exuberant phase transition phenomena.

\begin{center}
\textbf{Acknowledgements}
\end{center}

The authors dedicate this work to the memory of our dear friend, mentor and colleague F. G. Brady Moreira, who provided helpful suggestions on the early stages of this project.
We thank Paulo R. A. Campos for a critical reading of the manuscript.
We acknowledge financial support from UPE (PFA2019, PIAEXT2019) and the funding agencies FACEPE (APQ-0565-1.05/14, APQ-0707-1.05/14), CAPES and CNPq (167597/2017, 309961/2017, 436859/2018). 
The Boston University work was supported by NSF Grant PHY-1505000. 



\begin{center} 
\textbf{ADDITIONAL INFORMATION}
\end{center}

{\textbf{Author Contribution statement}}

\noindent 
ALMV, LFCP, LD and LRS conceived the project;
ALMV and LD performed the simulations and prepared the figures; 
ALMV, LFCP, LD, LRS and HES analyzed the results;
ALMV and LFCP wrote the manuscript; 
all authors contributed to the final version of this work.

{\textbf{Competing Financial Interests statement}}

\noindent
The authors declare no competing financial and non-financial interests.

\end{document}